\documentclass[lettersize,journal]{IEEEtran}
\usepackage{amsmath,amsfonts}
\usepackage{algorithmic}
\usepackage{algorithm}
\usepackage{array}
\usepackage[caption=false,font=normalsize,labelfont=sf,textfont=sf]{subfig}
\usepackage{textcomp}
\usepackage{stfloats}
\usepackage{url}
\usepackage{verbatim}
\usepackage{graphicx}
\usepackage{cite}
\usepackage{siunitx}
\usepackage{xcolor}
\usepackage{xspace}
\usepackage{ulem}

\begin{document}

\title{Lithographic integration of TES microcalorimeters with SQUID multiplexer circuits for large format spectrometers}

\author{\IEEEauthorblockN{Robinjeet Singh, Avirup Roy, Daniel Becker, Johnathan D. Gard, Mark W. Keller, John A. B. Mates, Kelsey M. Morgan,  Nathan J. Ortiz, Daniel R. Schmidt, Daniel S. Swetz, Joel N. Ullom, Leila R. Vale, Michael Vissers,  Galen C. O'Neil, Joel C. Weber}

\thanks{Authors thank John Biesecker, Sandra Diez, and Shannon M. Duff of National Institute of Standards and Technology, Boulder, CO, USA for useful discussions.}
\thanks{Robinjeet Singh, Avirup Roy, Daniel Becker, Joel N. Ullom, and Joel C. Weber are with the University of Colorado, Boulder, CO, USA and the Quantum Sensors Division of the National Institute of Standards and Technology, Boulder, CO, USA. Johnathan D. Gard is with the University of Colorado, Boulder, CO, USA.}
\thanks{Mark W. Keller, John A. B. Mates, Kelsey M. Morgan, Galen C O'Neil, Nathan J. Ortiz, Daniel R. Schmidt, Daniel S. Swetz, Leila R. Vale, and Michael Vissers  are with the Quantum Sensors Division of the National Institute of Standards and Technology, Boulder, CO, USA.}

}
            

\markboth{Journal of \LaTeX\ Class Files,~Vol.~14, No.~8, August~2021}%
{Shell \MakeLowercase{\textit{et al.}}: A Sample Article Using IEEEtran.cls for IEEE Journals}


\maketitle

\begin{abstract}

Arrays of hundreds or thousands of low temperature detectors have been deployed for many experiments, both bolometers for long wavelength applications and calorimeters for shorter wavelength applications. One challenge that is common to many of these arrays is the efficient use of focal plane area to achieve a large fill fraction of absorbers coupled to detectors. 
We are developing an integrated fabrication of soft X-ray transition edge sensors (TES) and microwave SQUID multiplexers ($\mu$MUX) with the goal of maximizing the fill fraction of the focal plane area on a scale of many thousand pixel detectors. We will utilize lithographically defined high density interconnects to circumvent limitations in existing solutions that use wirebonds or flip-chip bonds.  Here we report the first demonstration of combining TES and $\mu$MUX processes into a single TES-System-on-a-Chip (TES-SoC) fabrication on a silicon wafer. The $\mu$MUX SQUIDs and TES electrothermal feedback circuits are microfabricated first and protected with passivating SiO$_2$, then the TES devices and TES-to-SQUID interconnects are fabricated, and finally the protective layer is removed before the fabrication of the microwave resonators. We show that the microwave SQUIDs are functional and have reasonable yield, and that we are able to read out the transition temperature of the connected TESs using those SQUIDs.

\end{abstract}

\section{Introduction}
X-ray spectrometers that utilize transition edge sensor (TES) microcalorimeter arrays offer an ideal combination of exceptional energy resolution and quantum efficiency. These spectrometers are well positioned to revolutionize multiple science applications including nondestructive nanoscale  tomography \cite{Szypryt3KPixelDesign2021}, testing quantum electrodynamics using exotic atoms  \cite{OkumuraPRL2023}, ultrafast photochemistry \cite{Fowler_potential_2023}, and soft X-ray spectroscopy at X-ray light sources \cite{LeeRevScIns2019}. Towards this goal, TES spectrometers are now deployed to various synchrotron beamlines including the National Synchrotron Light Source (I and II), the Advanced Photon Source, the Stanford Synchrotron Radiation Lightsource, SPring-8, and BESSY II.

NIST's largest scale X-ray spectrometer utilizes a collection of microsnout submodules \cite{Szypryt3KPixelDesign2021}\cite{TOMCAT2023TAS}, with each microsnout containing an array of 250 TES microcalorimeter detectors. Each microsnout also contains four interface (IF) chips with bias and filtering circuits and four $\mu$MUX chips \cite{MatesAPL2017}. In order to maximize the fill-factor for active detector area, the IF and readout electronics chips are mounted ninety degrees out of the detector plane. The TES detectors are connected to their individual readout channels via around-the-corner wire bonds. The signal from the $\mu$MUX chips on multiple microsnouts can be read by one pair of input-output microwave cables. 

While $\mu$MUX technology has already enabled the current generation of compact and energy-efficient X-ray spectrometers with $\sim$ 1000 pixels, further improvement in TES pixel density is desirable for the next generation of spectrometers. One prime application that will benefit from the enhanced TES pixel density is carbon catalysis using Resonant Inelastic X-ray Scattering (RIXS) where currently deployed conventional spectrometers require collection times of tens of minutes to an hour. To reduce RIXS collection time to a few seconds, we are developing high-resolution X-ray spectrometers with $\sim$ 10,000 TES pixels. Here we present an integrated fabrication of microwave SQUID multiplexers and soft x-ray TESs on the same wafer that provides a path towards high pixel density in these spectrometers.

As mentioned, our current X-ray spectrometer technology utilizes wirebonds to interconnect TES to their $\mu$MUX based readout channels. The TES-to-SQUID wirebond pad pitch is 275 $\mu$m and sets the upper limit on the achievable pixel density on each microsnout. In addition, interconnecting 10,000 TES pixels to their readout electronics would require the  challenging completion of over 40,000 wirebonds. To address  these limitations, we have developed a new TES-System-on-Chip (TES-SoC) architecture where TES detectors are integrated with their $\mu$MUX readout electronics on a single monolithic wafer platform. We utilized lithographically defined TES-to-SQUID interconnects to eliminate the upper bound on the TES pixel density that is conventionally set by the wirebond pad pitch. The final design of our 10,000 pixel X-ray spectrometer will leverage TESs with overhanging absorbers to maximize photon collecting area \cite{OverhangingNasa2006}\cite{HaysOverAbs2016}\cite{wassell_Overhanging_2024}. We will combine TES-SoC with our previously demonstrated SOI-Flex technology \cite{o2024flexible} to fabricate TES-to-SQUID interconnects with the bias and readout components ninety degrees out of the detector focal plane. In this transaction, we will discuss our first demonstration of TES-SoC fabricated on bulk silicon where we studied effects of integration of TES detectors with $\mu$MUX read-out electronics.

In the following sections, we will describe the microfabrication methods for the TES-SoC including in-process wafer level diagnostics that aid in detecting  any drift in the critical current (I$_c$) of the $\mu$MUX circuits and tuning the superconducting critical temperature (T$_c$) of the TES detectors. We will then present measurements of a complete TES-SoC integrated chip, including resonator quality factor, bandwidth, and response to magnetic flux.

\begin{figure*}[!t]
\centering
\includegraphics[width=\textwidth]{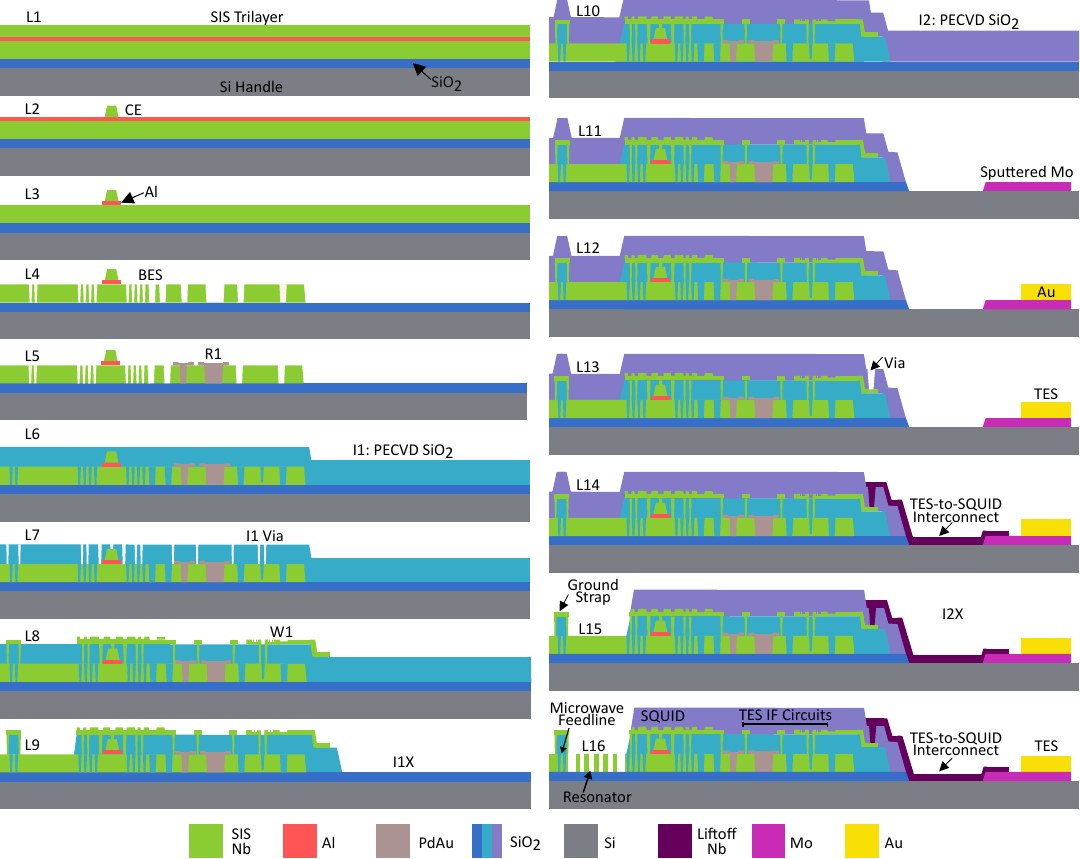}
 \caption{Layer schematic (not to scale) showing the microfabrication steps for TES-SoC on bulk Si. First, Nb-Al/Al$_x$O$_y$-Nb SIS trilayer is grown as L1. L2-L4 are etch steps for Nb counter-electrode (CE), aluminum (Al), and Nb base-electrode (BE) layers of SIS trilayers, respectively. Pd$_{0.53}$Au$_{0.47}$ shunt resistors are deposited as L5 using liftoff process. L6 and L7 include deposition of passivating PECVD SiO$_2$ (I1) and etching into I1 to open vias. Subsequent L8 layer is the deposition and etching of Nb (W1) that forms interconnects and termination wiring circuits using I1 vias. For layers L9 and L10, first I1 is removed in areas where microwave resonators, TES detectors, and TES-to-SQUID interconnects will be fabricated, then another layer of passivating SiO$_2$ (I2) is deposited to protect $\mu$MUX electronics. In L11, I2 is removed from areas where TES detectors and TES-to-SQUID interconnects will be fabricated, then a thin layer of molybdenum (Mo) is sputter-deposited, patterned, and etched using wet chemical bath. L12 is e-beam liftoff deposition of gold (Au) completing our MoAu bilayer TES detectors. L13 and L14 include RIE etching of vias into I2 and deposition of liftoff Nb to form TES-to-SQUID wiring interconnects. As final steps (L15 and L16), I2 and BE niobium are etched to form microwave resonators. The wafer is then diced into individual dies (step not shown in this schematic)} 

\label{fig_1}
\end{figure*}

\section{Fabrication Details}

Our $\mathrm{\mu}$MUX technology is based on niobium, aluminum, aluminum oxide, and niobium (Nb-Al/Al$_x$O$_y$-Nb) Superconductor-Insulator-Superconductor (SIS) tunnel Josephson junctions (JJ) \cite{GurvitchAPL1983}. We typically fabricate these $\mu$MUX devices on a 76.2 mm intrinsic undoped Si wafer grown with a float-zone (FZ) process. We first grow a very thin layer of thermal SiO$_2$ on top of the silicon wafer that acts as an etch stop for subsequent Nb processing.

We typically fabricate molybdenum-gold (MoAu) bilayer based TES microcalorimeter detectors \cite{weber2020development} on a Si wafer with a 500 nanometer thick layer of super-low stress low-pressure-chemical-vapor-deposition (LPCVD) silicon nitride (SiN$_x$). The Si handle underneath the TES electronics is selectively etched to form a SiN$_x$ membrane. Suspending the TES detectors on a SiN$_x$ membrane provides the thermal isolation required to set the time constant for the detector response. 

The need for two different thicknesses of varied dielectric materials for TES and $\mu$MUX devices, respectively, poses a challenge for integrated TES-SoC fabrication. To circumvent this challenge, we utilize a silicon on insulator (SOI) wafer platform for TES-SoC. As discussed for SOI-Flex in our previous work \cite{o2024flexible}, we chose a 4 $\mu$m thick silicon device layer and a thin (200 nm) buried oxide layer. The relatively thick device layer minimizes additional loss to the $\mu$MUX electronics due to the buried oxide layer. A 25 nm thick layer of thermal oxide is still grown on the SOI wafer to match the existing $\mu$MUX process. To thermally isolate the TES detectors in TES-SoC on SOI, our final design will suspend each detector on a membrane made from the silicon device layer instead of SiN$_x$. The Si handle is selectively etched underneath the desired area to release the Si membrane using buried oxide as an etch stop. Further details of the silicon membrane design are discussed in A. Roy et al. work being published concurrently \cite{ARoy2025}. 

For the initial demonstration described in this transaction, we fabricated a TES-SoC on a bulk intrinsic silicon with 32 TES detectors lithographically interconnected to their bias and $\mu$MUX readout components. The TES detectors being directly on thick silicon are not thermally suspended. With this demonstration, we tested the TES-SoC fabrication process, ensured proper working of all components at cryogenic temperatures, and demonstrated the first lithographic integration of TES detectors with $\mu$MUX readout.

\subsection{Readout Fabrication}

The complete fabrication flow for TES-SoC on bulk silicon is depicted in Fig. \ref{fig_1}. We utilize both numeric and descriptive layer IDs. We start with a 76.2 mm (3 inch) diameter FZ silicon wafer with 25 nm of thermal SiO$_2$. In L1, we deposit a Nb-Al/Al$_x$O$_y$-Nb SIS tri-layer on the whole wafer using sputter deposition and diffusive oxidation techniques. We first deposit 200 nm of niobium as base-electrode (BE) followed by $\sim$ 15 nm of aluminum (Al). The Al thickness is selected to minimize any pinhole defects that may lead to shorted junctions. Once BE and Al are deposited, the wafer is transferred without breaking vacuum into an oxidation chamber to form an insulating Al$_x$O$_y$ layer. We adjust the depth of oxidation and hence the critical current density (J$_c$) of our SIS junctions using Eq. \ref{eq1}.

\begin{align}
    J_c &\approx \eta{E^{-0.51}}, \label{eq1} \\
    E &= P \times t \nonumber,
\end{align}
 where the pre-factor $\eta$ is specific to our oxidation chamber and the oxidation exposure, $E$, is the product of the oxygen pressure, $P$, and the oxidation duration, $t$. For $\mu$MUX devices, P is set to 1200 Pa (9 Torr) and we achieve J$_c \approx 0.8$ $\mathrm{{\mu}A/\mu{m^2}}$ with a duration of 312 min. This corresponds to an estimated thickness of Al$_x$O$_y$ between 1 nm and 2 nm, calculated using WKB approximation for a rectangular-barrier tunneling model \cite{Simmons1963}. 

After the oxidation of Al, again without breaking the vacuum, we move the wafer back into the process chamber for deposition of 120 nm of niobium counter-electrode (CE), completing our SIS tri-layer. We want the room temperature residual stress of the SIS tri-layer to be overall compressive to compensate for the change as metal layers contract at cryogenic temperatures \cite{Imamura1992}.  We deposit Al at a set chamber pressure of 0.67 Pascal and adjust the chamber pressure for the niobium BE and CE layers to yield the desired film stress.

After L1, we lithographically define the majority of our $\mu${MUX} circuits that were first detailed in \cite{bmatesthesis}. These circuits consist of dissipationless gradiometric rf-SQUIDs, each connected to a quarter-wave microwave resonator, and filter circuits for the flux ramp and TES bias. The output of the microwave resonators is capacitively coupled to a common coplanar microwave feedline, with the coupling capacitance set by changing dimensions of the interdigitated pattern. In L2, we pattern and etch CE using step and repeat i-line optical lithography and a sloped etch using O$_2$:SF$_6$ in a parallel plate reactive-ion-etch (RIE) system. The designed area of CE squares is 6.25 $\mu$m$^2$ targeting the desired I$_c$ of 5 $\mu$A for our JJs. We use an oxygen plasma ash and solvent-based clean after each etch step to prepare the wafer for subsequent processing. 

In L3, we pattern aluminum pillars that are etched using tetramethylammonium hydroxide (TMAH) based wet chemistry. The Al features are made wider than CE to avoid any short circuit of CE to BE due to aluminum undercut. L4 involves sloped etching of BE (BES) to define the JJ base-electrode, slotted washers for SQUID inductors, microwave feedline, and input filter circuits. For TES-SoC, BES also defines the interdigitated circuits for the TES bias shunt resistors and the Nyquist inductor filters. We use O$_2$:CF$_4$ based RIE chemistry that ensures a sloped etch profile to facilitate conformal coverage of the subsequent metal layers. We note that the coplanar waveguides defining the quarter-wave microwave resonators are designed to have vertical sidewalls and are etched as a separate processing step.

\begin{figure}
\centering
\includegraphics[width=0.49\textwidth]{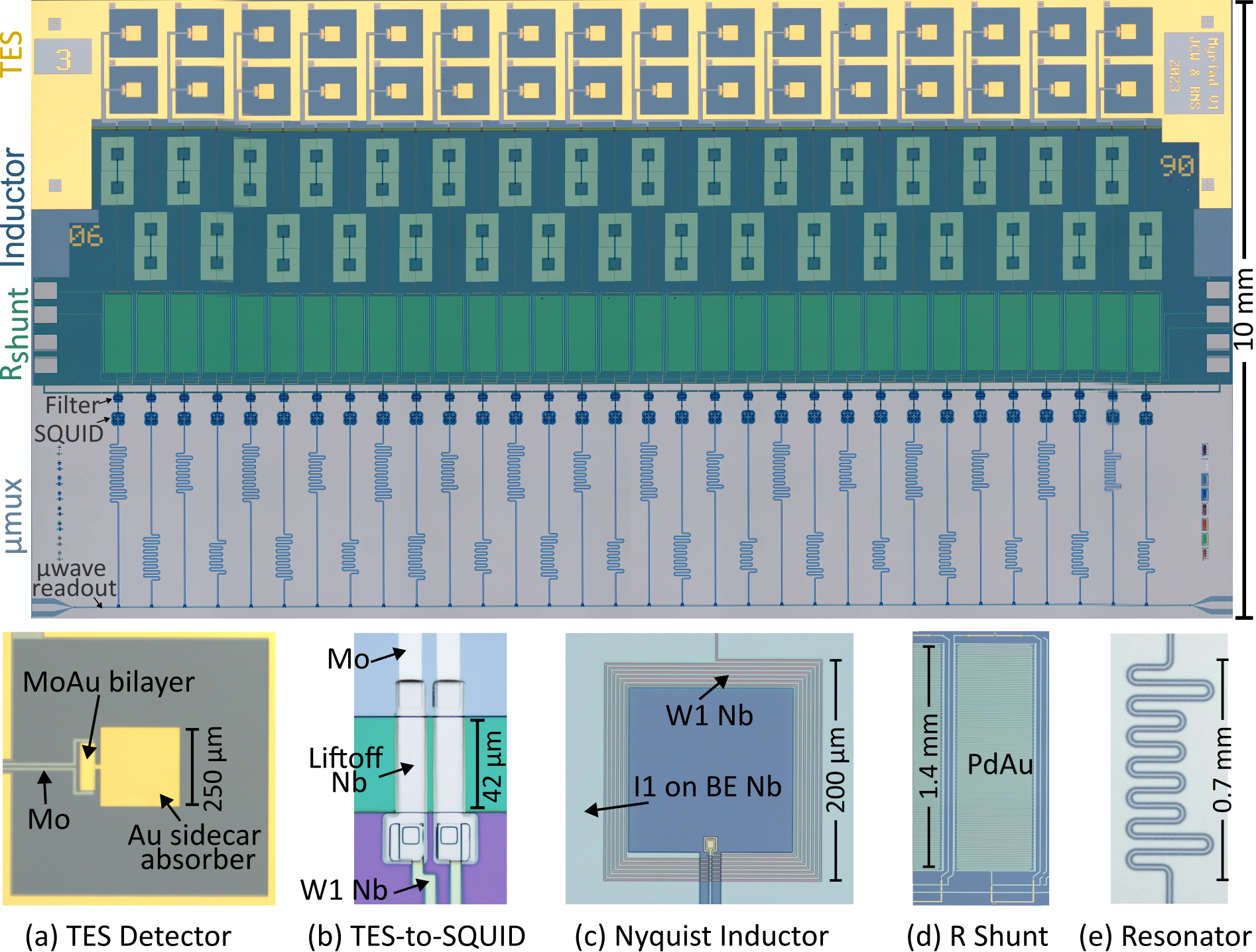}
\caption{Optical micrograph of a TES-SoC die showing MoAu bilayer based TES detectors, IF electronics, and $\mu$MUX readout integrated onto a single monolithic wafer platform. The chip dimension are 10 mm $\times$ 20 mm. TES detectors are fabricated directly on bulk silicon and are not suspended. (a), (b), (c), (d), and (e) are higher magnification optical micrographs of the TES detectors, TES-to-SQUID liftoff niobium interconnects, Nyquist inductors, shunt resistors, and the $\mu$MUX resonators, respectively.}
\label{TES-SoC}
\end{figure}

In subsequent L5, we deposit 135 nm of Pd$_{0.53}$Au$_{0.47}$ as R1 to shunt the inductive input filters for the $\mu$MUX SQUIDs and bias resistors for the TES detectors. We use liftoff photolithography and e-beam evaporation techniques. Thin layers of titanium are used before and after R1 to promote adhesion. Excess metal is removed using organic solvents and ultrasonic agitation. After R1, we deposit a capping layer of 350 nm thick plasma enhanced chemical vapor deposition (PECVD) SiO$_2$ (I1) as L6 to protect the SIS metals from the top niobium wiring (W1). For L7, we pattern and etch via holes into I1 (using O2:CHF3 RIE chemistry) to selectively interconnect underlying SIS layers using the W1 wiring layer, which consists of 400 nm of sputtered niobium. W1 is patterned and etched using O$_2$:CF$_4$ based RIE chemistry to complete L8. W1 layer forms CE-to-BE interconnects for our JJ, grounding straps over the microwave feedline to prevent waveguide slot-mode excitations, and complete input flux ramp and TES bias lines \cite{bmatesthesis}. W1 also forms the inductor washers for the TES IF circuits.

\subsection{JJ and SQUID Characterization}
To monitor the microfabrication processes, we employ various in-process wafer-level diagnostics on test chips that are strategically distributed throughout the wafer. As a diagnostic for our $\mu$MUX JJs, we track the room temperature resistance R$_n$ of 200 test JJs in series and utilize the Ambegaokar-Baratoff approximation \cite{AMBPRL} to estimate I$_c$ from R$_n$,  

\begin{align}
    I_cR_n \approx {\pi{\Delta}\over2e},
    \label{eq2}
\end{align}
where $e$ is the elementary charge and $\Delta$ is the superconducting energy gap calculated from BCS theory. As shown in our previous work \cite{jones_qualification_2024}, the R$_n$ measurements provide a reliable in-fab diagnostic to ensure our JJ performance is within the design threshold. For post-fabrication, cryogenic testing of our rf-SQUIDs, we track the quantity $\lambda$ defined as the ratio of self-inductance (L$_s$) and Josephson-inductance (L$_J$) within the SQUID. For dissipationless rf-SQUIDs to respond monotonically to the applied flux, the $\lambda$ is desired to be $\sim$ 0.33 $\pm$ 20$\%$. The L$_s$ of our $\mu$MUX SQUIDs is designed to be 22.4 pH,  setting the target per unit area R$_n$ to be (68.8 $\pm$ 14) $\Omega/\mu$m$^2$. The uncertainty accounts for wafer-to-wafer microfabrication variations.

In TES-SoC, we tracked the performance of test JJ arrays distributed at the center and corners of the wafer. Based on these measurements, we calculated the average R$_n$ of TES-SoC JJs with area 6.25 $\mu$m$^2$ to be $\sim$ 362 $\Omega$. This equates to per unit area room temperature resistance of $\sim$ 58 $\Omega/\mu$m$^2$, well within uncertainty due to variations in microfabrication. 

\subsection{TES Detectors Fabrication}

Once the room temperature diagnostic of our JJ is confirmed to be within specification, we proceed to L9 where we etch the I1 at sites where TES MoAu bilayers, TES-to-SQUID interconnects, and microwave resonators will be formed. In L10, to protect the $\mu$MUX and IF circuits from degradation during MoAu processing, we deposit 400 nm of PECVD SiO$_2$ (I2). We then proceed to microfabricate MoAu bilayer TES detectors as described in \cite{weber2020development}. 
To define TES detectors, we follow the hybrid additive-subtractive TES process where 45 nm of molybdenum is sputter-deposited, patterned, and then etched using a phosphoric acid and acetic acid-based commercial etchant solution. In L12, a 411 nm thick layer of gold is deposited on top of molybdenum features via e-beam evaporation and liftoff process after a 1 min Ar RF plasma clean to improve MoAu interface transparency. The whole wafer is then baked for 20 min in an ambient environment at 150 $^{\circ}$C to anneal the MoAu bilayer and stabilize TES T$_c$ and R$_n$ from additional thermal processing. In order to accurately tune the target T$_c$ of the fabricated TES, we perform wafer-level cryogenic T$_c$ measurements as described in work by J. Weber et. al published concurrently with this transaction \cite{WeberLTD2025}. Additional gold is added or removed based on our T$_c$ diagnostics to obtain the final target T$_c$. A thinner layer of 250 nm gold with a 2 nm Ti adhesion layer is then deposited as a separate e-beam evaporation and lift-off process to define the absorbers for our TES detectors. A final ambient environment anneal at 150 $^{\circ}$C for 20 min is completed to ensure good thermal contact between all Au layers.

We note that in our current demonstration of TES-SoC, we removed both I2 and the thin thermal oxide from TES detector sites. MoAu bilayers were then directly deposited on top of the silicon substrate. We observed across-wafer T$_c$ non-uniformities and broad superconducting-normal transitions in our TES detectors. We attribute these negative characteristics to molybdenum-silicide formation. In future designs, we will fabricate TES on top of I2 SiO$_2$.

\subsection{TES-to-SQUID Interconnects}

After the MoAu bilayer process, in L13, we etch via holes into the I2 layer to expose W1 for TES-to-SQUID niobium interconnects defined by L14 where 500 nm of sputtered Nb is deposited by lift-off process. During this step, the detectors and other components are protected by photoresist, whereas doing this step with a niobium etch process would require an additional passivating SiO$_2$ layer that would complicate TES-SoC integration.

\subsection{Microwave Resonator Fabrication}

Once the MoAu bilayers, gold  absorbers, and TES-to-SQUID interconnects are fabricated, we check the R$_n$ of the test structures to track any unexpected changes due to additional processing. In L15, we pattern and etch the I2 layer to expose BE niobium at sites for microwave resonators that are patterned and etched in L16 using SF$_6$ chemistry. 

\subsection{TES-SoC Die Release}
In the current demonstration of TES-SoC on bulk Si, individual dies are released using mechanical dicing. Fig. \ref{TES-SoC} shows a micrograph of the diced TES-SoC die. The TES detectors, located on bulk silicon with no thermal isolation, were not expected to be suitable for X-ray spectrum or pulse measurements. The final design of TES-SoC on SOI will be released using deep reactive ion etch (DRIE) where the Si device layer will be first patterned and etched to define the chip boundary. We will then selectively etch the Si handle from the backside using DRIE to simultaneously suspend TES detectors on Si membranes and release the individual dies.

\section{Results}

After fabrication and singulation of individual TES-SoC on bulk Si dies, we carried out   cryogenic testing to confirm the final T$_c$ of the fabricated TES detectors, response of the rf-SQUIDs to the input flux, and the quality factors of the microwave resonators.

Since the TES detectors were fabricated directly on the silicon and were not suspended on thin SiN or Si membranes, their time constant was too fast and current too large to be read out well with these $\mu$MUX devices. We measured an average transition temperature of (141 $\pm$ 3) mK for several TES-SoC detectors by varying the bath temperature in small increments.

The I$_c$ of the TES-to-SQUID sputtered niobium was tracked on a separate test wafer and was measured to be much higher than the I$_c$ of the molybdenum for wiring traces of similar dimensions. We measured I$_c$ greater than 80 mA for liftoff niobium wiring traces of varying lengths and widths. The response of the TES detectors and their transition temperatures were measured on the integrated TES-SoC to confirm that TES-to-SQUID interconnects were functioning as designed.

\begin{figure}
\includegraphics[width=0.47\textwidth]{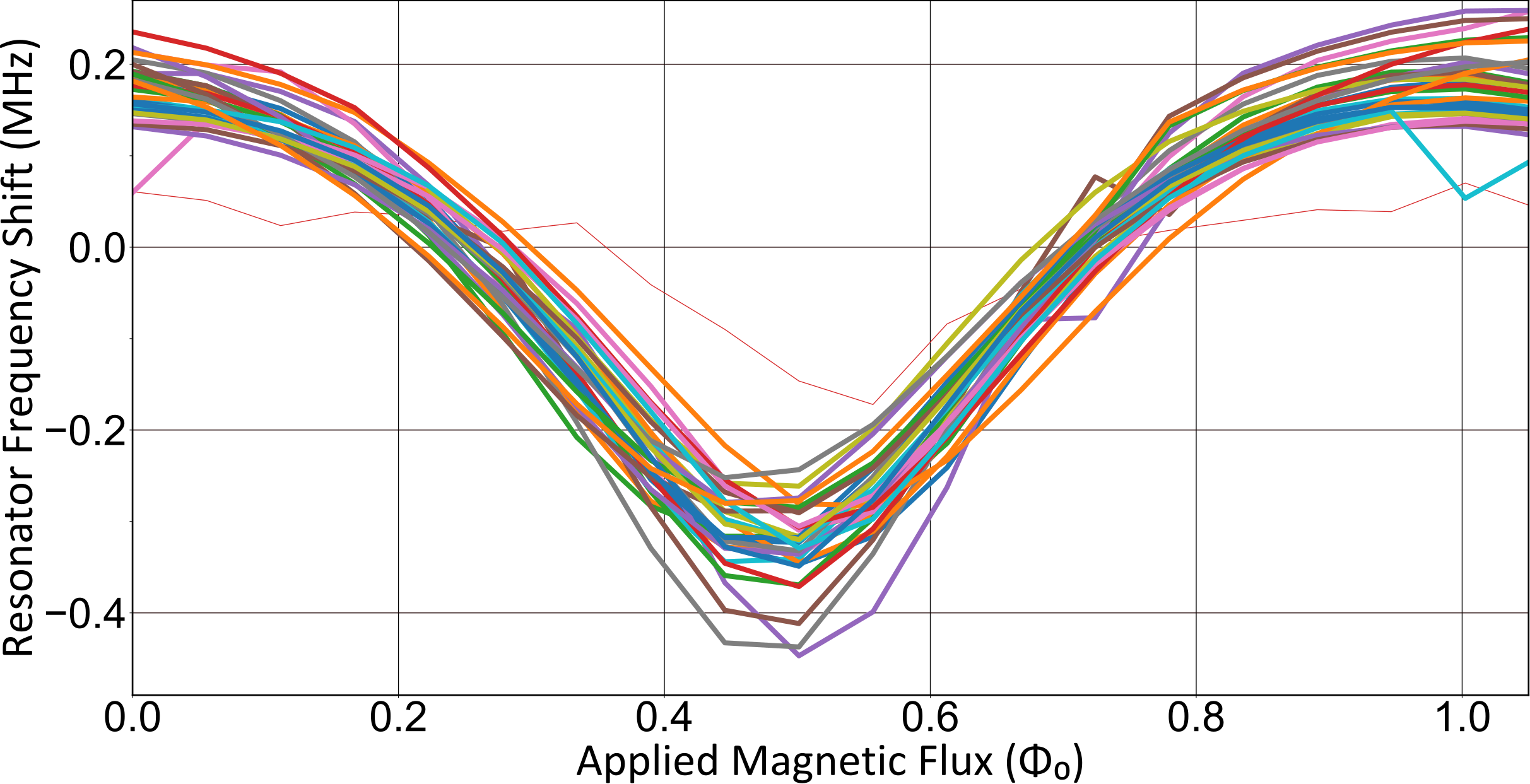}
\caption{Data showing shifts in the resonant frequencies of the 32 resonators on TES-SoC, as a function of applied magnetic flux. An x-axis offset is applied to match the phase between all channels. The average value of M$_{in}$ $\sim$ 9.2 pH is calculated by fitting the frequency shifts to theoretical models.}
\label{fig_frShift}
\end{figure}

We measured the key design parameters of the  $\mu$MUX devices and found that they matched the design values well and had a relatively high yield of above 96\%.  Fig. \ref{fig_frShift} shows shifts in the peak-to-peak frequencies of $\mu$MUX resonators as a function of applied magnetic flux. We observed an average of 0.43 MHz peak-to-peak frequency shifts in the resonators. We calculate the average mutual inductance between the input circuits and the rf-SQUIDs (M$_{in}$) and $\lambda$ by fitting the measurements shown in  Fig. \ref{fig_frShift} to theoretical models. M$_{in}$ for TES-SoC $\mu$MUX circuits is calculated to be $\sim$ 9.2 pH.  

\begin{figure}
\centering
\includegraphics[width=0.47\textwidth]{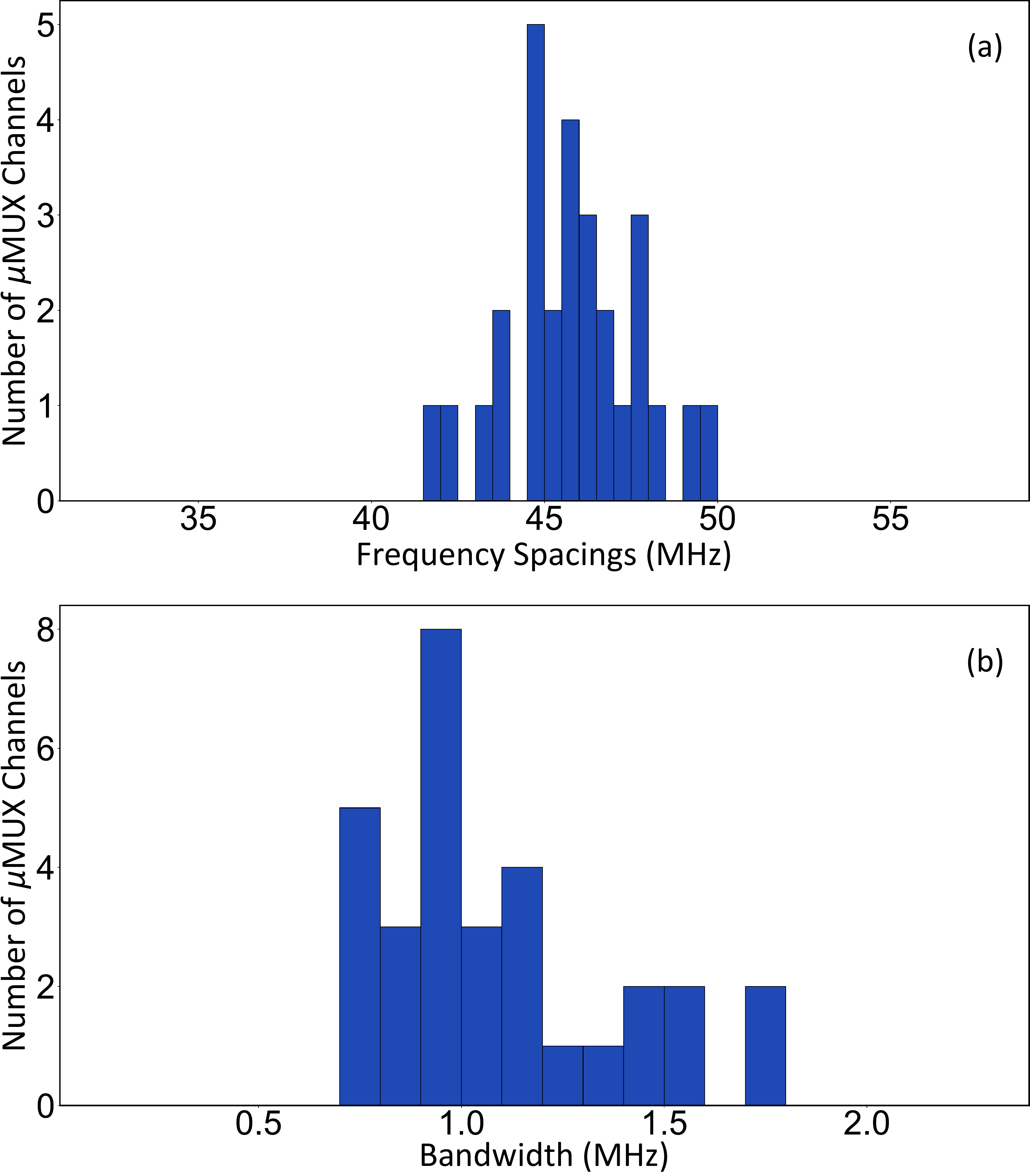}
\caption{Histograms showing (a) the frequency spacings between resonators, (b) the bandwidth of resonators for the $\mu$MUX channels on the TES-SoC.}
\label{bandwidth-freqspacing}
\end{figure}

The TES-SoC resonators were designed to have 1 MHz bandwidth and 45 MHz spacing between adjacent resonators. Figure \ref{bandwidth-freqspacing} shows histograms of (a) measured frequency spacings and (b) resonator bandwidths. Measured values match the target values with variations that are either by design or are within our fabrication uncertainty budget.   

\begin{figure}
\centering
\includegraphics[width=0.47\textwidth]{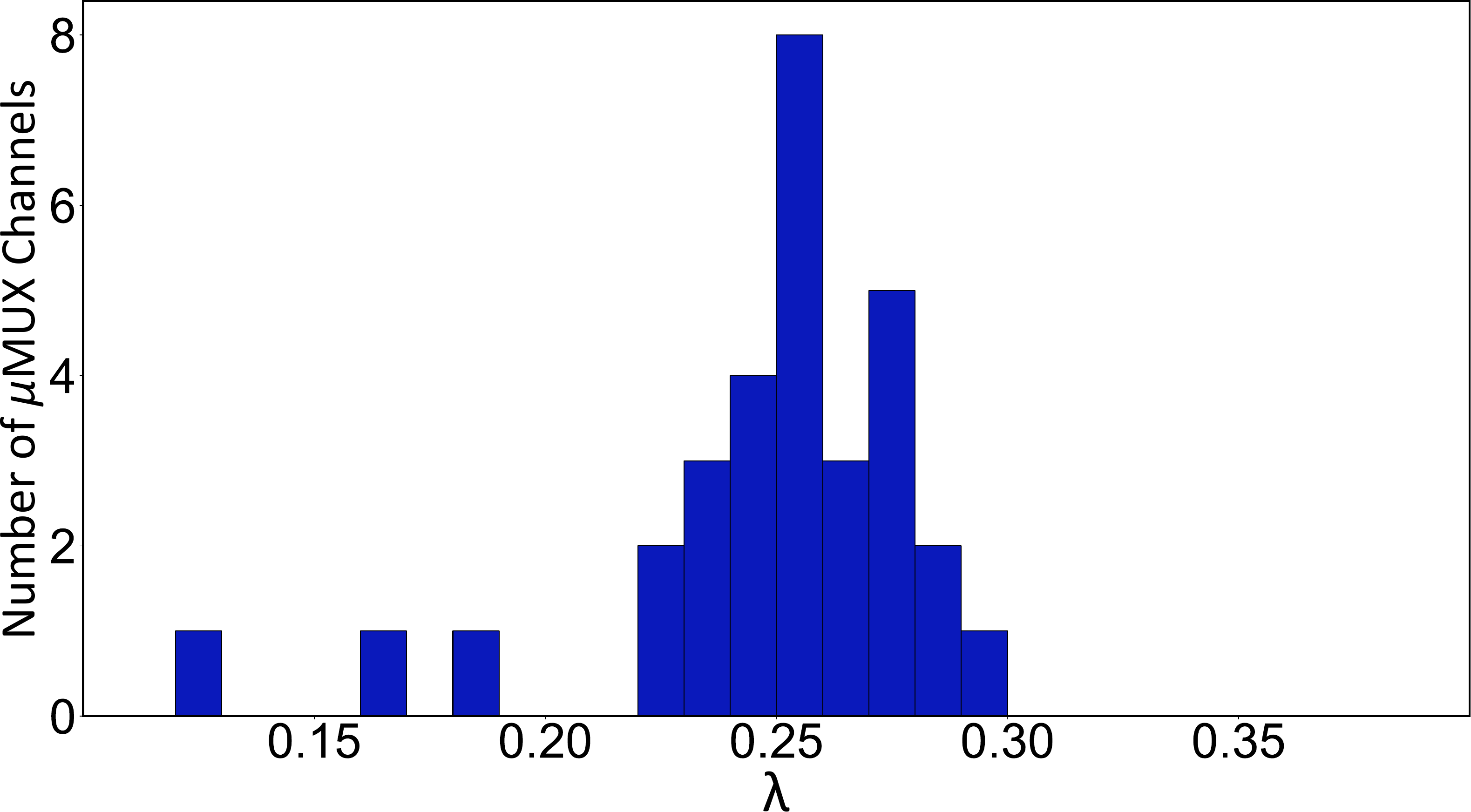}
\caption{Histogram showing $\lambda$ spread of all the $\mu$MUX channels on the TES-SoC.}
\label{fig_lambda}
\end{figure}

We calculated $\lambda$ based on modulation depth of the resonator frequencies due to applied magnetic flux. Figure \ref{fig_lambda} is a histogram with average $\lambda$ of $\sim$ 0.25 for all the measured $\mu$MUX channels. We note that although the $\lambda$ value is within design tolerance, it may have been slightly suppressed due to the microwave probe power being too high.

One crucial aspect in designing the $\mu$MUX die form factor is to select the die dimensions that will avoid interference from resonant standing waves or so-called box modes in the silicon substrate. The resonant frequencies of these box modes in a rectangular waveguide can be calculated using Eq. \ref{Eq_box} \cite{collin2007foundations}.

\begin{align}
    f_{mnp} = \frac{c}{2\sqrt{\epsilon_r\mu_r}}\sqrt{\left(\frac{n}{l}\right)^2+\left(\frac{m}{w}\right)^2 + \left(\frac{p}{h}\right)^2}
    \label{Eq_box}
\end{align}
where $m, n, p$ are the order of TEM modes, $l, w, h$ are the length, width, and height of the rectangular waveguide, $c$ is the speed of light in vacuum, and $\epsilon_r$ and $\mu_r$ are the relative permittivity and TEM mode permeability of the waveguide material, respectively. 

\begin{figure}
\centering
\includegraphics[width=0.47\textwidth]{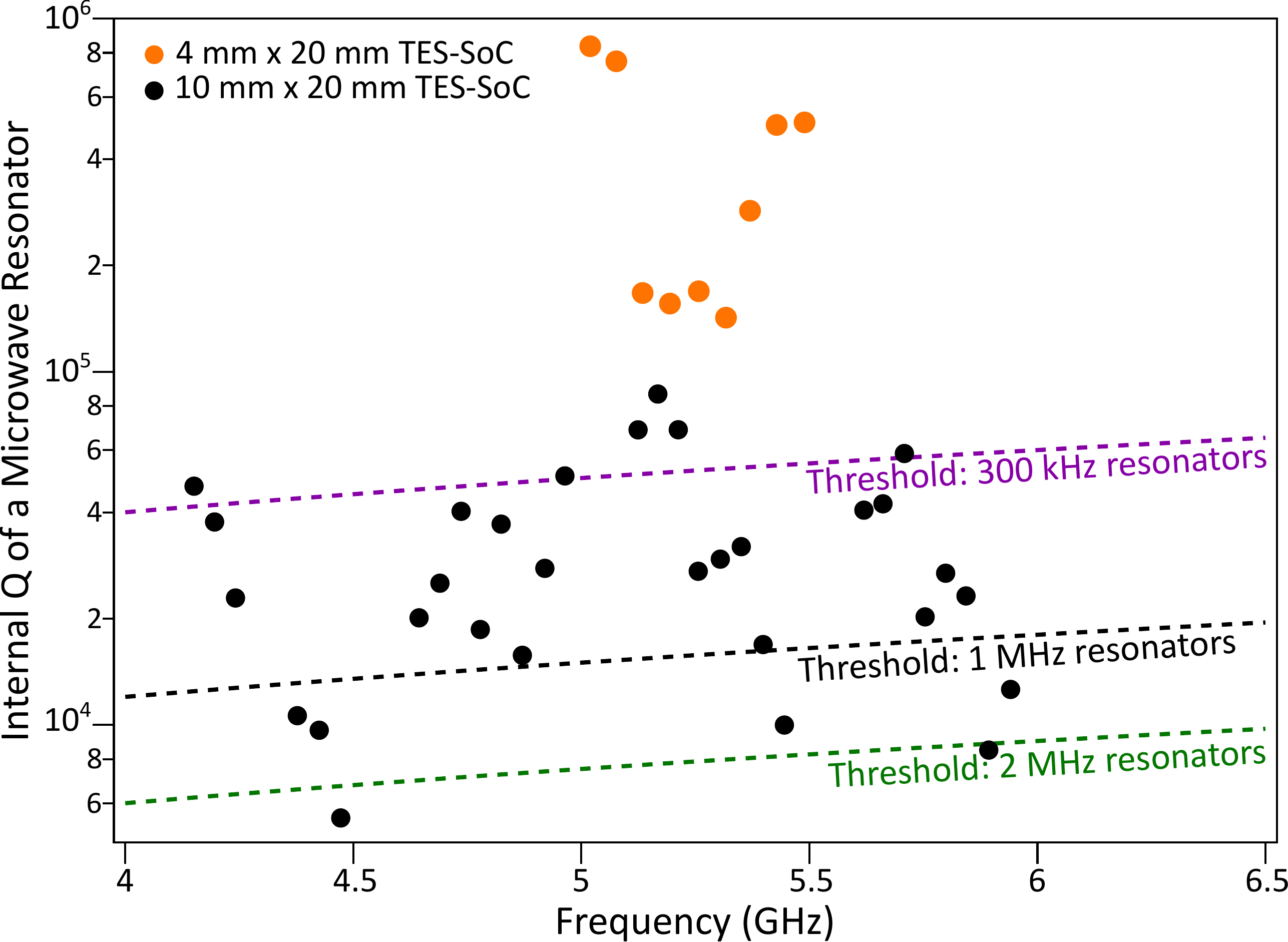}
\caption{Internal Quality factor (Q$_i$) of the TES-SoC resonators measured on two chips with different dimensions. The test TES-SoC chip is $4~mm \times 20~mm$ similar to a standalone $\mu$MUX die. The TES-SoC with fully integrated TES detectors, IF and $\mu$MUX circuits is $10~mm \times 20~mm$. We attribute degraded Q$_i$ of resonators on the the fully integrated TES-SoC to interference from the standing-wave box modes in the silicon substrate. Threshold traces show the minimum Q$_i$ for each bandwidth of the $\mu$MUX resonators.}

\label{QiTESSOC}
\end{figure}

For a traditional standalone $\mu$MUX chip, lateral dimensions of 4 mm $\times$ 20 mm are selected to place the box-mode frequencies outside the (4 to 8) GHz range of interest.  The TES-SoC chip described in this work is larger, 10 mm x 20 mm, and we expect increased loss of the microwave resonators due to interference with multiple box-modes. Figure \ref{QiTESSOC} shows the measured internal quality factor (Q$_i$) of resonators on the 10~mm x 20~mm TES-SoC chip which is expected to have box mode degradation. In the same figure, resonators on a 4~mm $\times$ 20~mm test chip, fabricated simultaneously on the TES-SoC wafer, are shown to have much larger Q$_i$ due to reduced interference from the box-modes.  Threshold Q$_i$ for 300 kHz, 1 MHz, and 2 MHz resonators are also shown in Fig. \ref{QiTESSOC}; internal quality factors lower than the threshold will cause poor noise performance. We note that even with box mode degradation, 1 MHz $\mu$MUX threshold was surpassed by above 80\% of the resonators on 10~mm $\times$ 20~mm TES-SoC. Above 90\% of the TES-SoC resonators surpassed 2 MHz threshold. 

\begin{figure}
\centering
\includegraphics[width=0.47\textwidth]{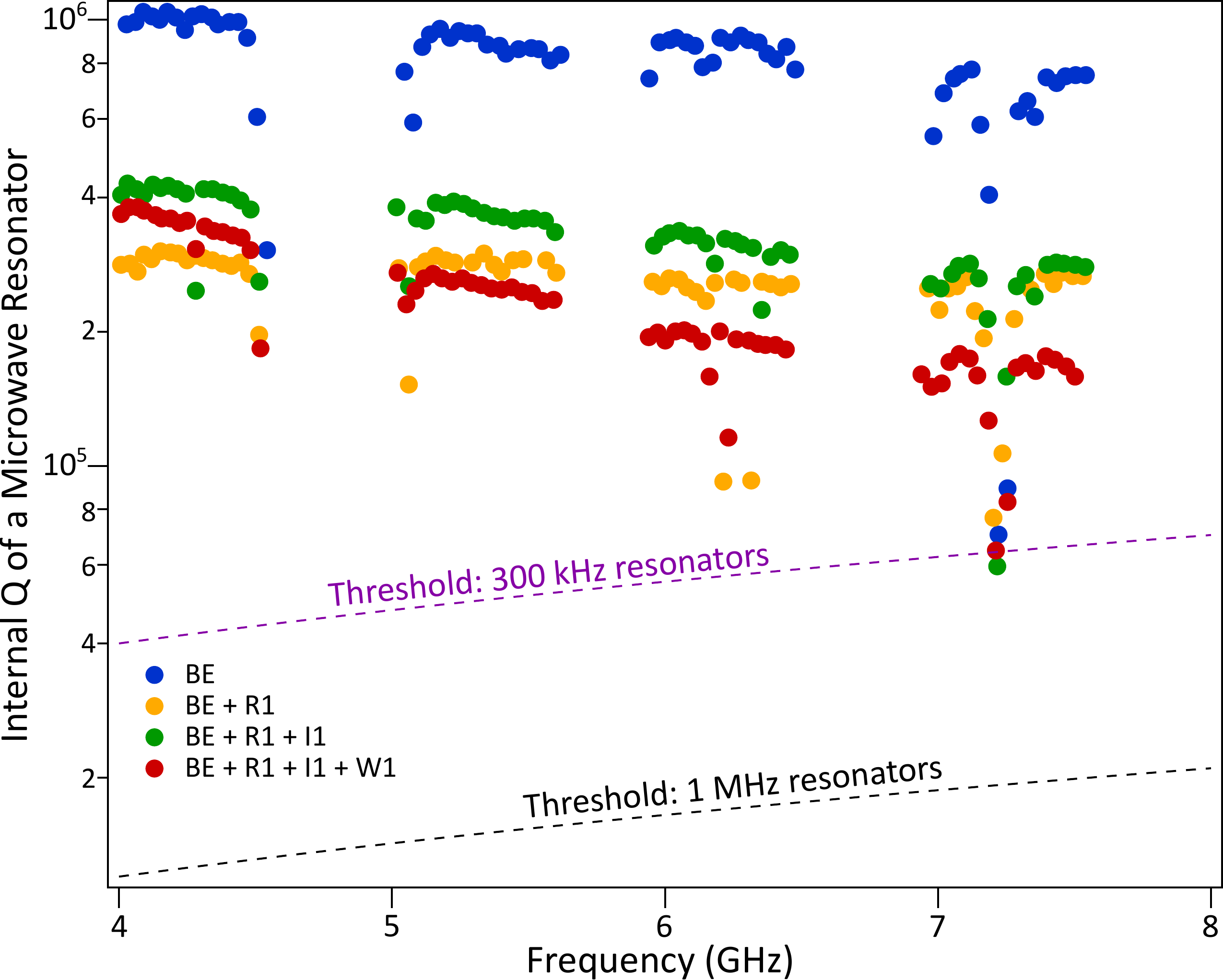}
\caption{Q$_i$ of bare microwave resonators after different amount of processing. The wafer with only a single layer of resonators had much higher internal quality factor than any of the other wafers.  While additional processing degrade the Q$_i$ of our microwave resonators, the exact mechanism causing these degradations is still a topic of active research. Reduced Q$_i$ of resonators at $\sim$ 7.2 GHz is possibly due to interference with cavity modes of the chip enclosure. Threshold traces are drawn for a qualitative comparison to Q$_i$ of TES-SoC resonators.}
\label{QiTri}
\end{figure}

Figure \ref{QiTri} shows process dependence of Q$_i$ in our standard $\mu$MUX fabrication. Four different test wafers were fabricated starting with a 200~nm thick sputtered Nb on an oxidized intrinsic Si wafer. Microwave resonators were etched in the Nb film after each of the progressive processing steps: BE deposition, BE deposition + R1 liftoff, BE deposition + R1 liftoff + I1 deposition and etch, and BE deposition + R1 liftoff + I1 deposition and etch + W1 deposition and etch, respectively. Sixteen chips from four processed wafers were measured to study the degradations in Q$_i$ of the resonators. All of the resonators on 4~mm $\times$ 20~mm test TES-SoC shown in Fig. \ref{QiTESSOC} have Q$_i$ comparable to the standard $\mu$MUX resonators in Fig. \ref{QiTri} and are at least an order of magnitude larger than the threshold for 1~MHz resonators. We did not observe any evidence of additional degradation in Q$_i$ due to added microfabrication steps for TES-SoC processing.

In future designs of larger form-factor TES-SoC, we will implement grounding structures within the chip to effectively reduce the dimensions of the microwave cavity and minimize box-mode losses. Based on our observations and planned future design improvements, we estimate that the TES-SoC process will produce high yield $\mu$MUX resonators with various bandwidths including 1 MHz targeted for soft x-ray TES readout. This is a topic of ongoing research and is key to extending the number of $\mu$MUX channels on each TES-SoC die by orders of magnitude for a 10,000 TES pixel spectrometer.

Further statistics are needed to confidently pinpoint the particular microfabrication process causing the major reduction in the Q$_i$ and is a topic of ongoing investigation.

\section{Conclusion}
We have demonstrated the integration of TES microcalorimeter detectors with $\mu$MUX readout components using lithographically defined niobium wiring interconnects. To our knowledge, this is the first demonstration of such integration for superconducting low-temperature detectors. We note that in addition to enabling a new generation of high TES pixel density X-ray spectrometers, TES-SoC technology is useful for a variety of applications including integrated TES based Gamma-ray detectors and large array CMB telescopes. Our TES-SoC technology minimizes the need for chip-level packaging  traditionally required to interconnect TES detectors to their readout electronics through TES-to-SQUID wirebonding or flip-chip bonding. The TES-SoC process requires a large number of fabrication steps, but may still be cost-effective for many applications.  The lithographically defined wiring interconnect in our TES-SoC technology does not require any additional thermal processing and hence the integration process is not expected to degrade the performance of our TES detectors. The yield of the $\mu$MUX devices was above threshold and good.

\bibliographystyle{IEEEtran}
\bibliography{references}

\vfill

\end{document}